# Evaluation of the linear theory satisfiability limits in propagation of the internal gravity waves

**V.V.Bulatov, Y.V.Vladimirov**
**Institute for Problems in Mechanics**
**Russian Academy of Sciences**
**Pr.Vernadskogo 101-1, 117526 Moscow, Russia**
**bulatov@index-xx.ru**

*Abstract*

*A problem of the linear theory satisfaction limits in propagation of the internal gravity waves is considered. It is evident that internal gravity waves excitation, propagation in actual practice is highly nonlinear phenomenon. However with some reasonable assumptions it is possible to linearized equations of internal waves generation and propagation. It is shown that in interesting for us wavelength range we can use linear approximation during study of internal gravity waves dynamics. Similarly it is easy to evaluate also influence of other corrections to the linear theory of internal gravity waves generation and propagation, and the obtained results indicate adequacy and supportability of linear model wave dynamics.*

The present work will consider the matter of internal gravitational waves linear theory satisfiability limits as the treatise main results in [1-5] were obtained in this approximation. It is evident that internal gravity waves excitation, propagation in actual practice is highly nonlinear phenomenon [6-9]. However with some reasonable assumptions it is possible to linearize equations of internal waves generation and propagation. In [1-5] mainly linear approximations for internal gravitational wave were studied at big distances from the excitation source, in the case, when the source is replaced with a certain model, horizontal changes of the stratified medium primals were considered to be minor ones at the scales of wave length etc. Therefore evaluation of satisfiability limits made in [1-5] assumptions is of interest.

Combined hydrodynamic equations in "traditional" approximation for Coriolis forces resulting from the terrestrial rotation and in adiabatic approximation for the equitation of condition to be written as [6-9]

$$\rho \frac{dU_1}{dt} + \frac{\partial p}{\partial x} = -\rho f U_2 + \nu \left[ \Delta_3 U_1 + \frac{\partial}{\partial x} div \vec{U} \right] + F_x$$

$$\rho \frac{dU_2}{dt} + \frac{\partial p}{\partial y} = -\rho f U_1 + \nu \left[ \Delta_3 U_2 + \frac{\partial}{\partial y} div \vec{U} \right] + F_y \qquad (1)$$

$$\rho \frac{dW}{dt} + \frac{\partial p}{\partial x} + g\rho = \nu \left[ \Delta_3 W + \frac{\partial}{\partial z} div \vec{U} \right] + F_z$$

$$\frac{1}{c^2} \frac{\partial p}{\partial t} = \frac{\partial \rho}{\partial t}$$

$$\frac{\partial p}{\partial t} + div\, \rho \vec{U} = M$$

$$\frac{d}{dt} = \frac{\partial}{\partial t} + U_1 \frac{\partial}{\partial x} + U_2 \frac{\partial}{\partial y} + W \frac{\partial}{\partial z}; \quad f = 2\Omega \sin \theta$$

$$\Delta = \frac{\partial^2}{\partial x^2} + \frac{\partial^2}{\partial y^2}, \quad \Delta_3 = \Delta + \frac{\partial^2}{\partial z^2}$$

where $\Omega = 7.27 \cdot 10^{-5}$ sec$^{-1}$ is the terrestrial rotation frequency, $\theta$ is the geographical latitude, $\nu = 10^{-2}$ cm$^2$/sec is the viscosity ratio; are the velocity components, $p, \rho$ are pressure and density, axis $z$ is up-directed, $g = 980$ cm/sec$^2$ is the acceleration of gravity, $c = 1.5 \cdot 10^5$ cm/sec$^2$ is the sound speed; at last, $F_x, F_y, F_z$ are densities of volume forces acting on fluid and $M$ is the mass sources density.

Taking into account the fourth equation (1), the equation for $\frac{\partial p}{\partial t}$ can be written over as

$$\frac{1}{c^2} = \frac{\partial p}{\partial t} + \rho \, div \, \vec{U} = M$$

Further either vertically infinite medium or the layer restricted by the bottom $z=-H$ and the free surface $z = \varsigma(x,y,t)$ will be considered. With the bottom viscosity one should lay the condition for adhesion of the mixture no-slip condition $U_1 = U_2 = W = 0$ with $z = -H(y,z)$, which would form the appropriate near-boundary layer. However particle s velocity small values of ocean internal gravitational waves are characteristic, in the region of 10 cm/sec or less and, accordingly, rate of shear small values. Therefore viscosity effects result in only small corrections to ideal fluid solution, in this way at the bottom it is possible to be restricted by the no-fluid-loss condition meaning that the vector of speed at the bottom must be tangent to surface.

$$W - U_1 \frac{\partial H}{\partial x} - U_2 \frac{\partial H}{\partial y} = 0 \quad (z = -H(x,y)) \tag{2A}$$

For horizontal bottom $(H = const)$ this condition is simplified:

$$W = 0 \quad (z = -H) \tag{2B}$$

On ocean free surface $z = \varsigma(x, y, z)$ tow boundary conditions are laid – kinematic and dynamic ones. The kinematic condition requires for fluid particles orthogonal to a surface velocity component $(U,V,W)$ coincides with surface drift velocity. This condition results in the congruence

$$W\big|_{z=\varsigma} = \frac{\partial \varsigma}{\partial t} + U_1 \frac{\partial \varsigma}{\partial x} + U_2 \frac{\partial \varsigma}{\partial y} = \frac{d\varsigma}{dt} \tag{3}$$

The dynamic condition requires the surface pressure coincides with the atmospheric pressure $p_a = (x, y, t)$:

$$p(x, y, \varsigma(x, y, t), t) = p_a(x, y, t) \tag{3A}$$

Thereafter atmospheric pressure variations waves won't be considered, therefore later we will consider $p_a = 0$, i.e.

$$p(x, y, \varsigma(x, y, t)) = 0 \qquad (3B)$$

We linearize the combined equations (1) and the boundary conditions (2)-(3) with respect to a standstill

$$U_1 = U_2 = W = 0; \quad \rho = \rho_0(z), \quad p = p_0(z) = -g\int_0^z \rho_0(z)dz$$

For this purpose let us assume $p = p_0 + \hat{p}, \quad \rho = \rho_0 + \hat{\rho}; \quad U_1 = \hat{U}_1, \quad U_2 = \hat{U}_2, \quad W = \hat{W}$ and let us write out equations for $\hat{p}, \hat{\rho}, \hat{U}_1, \hat{U}_2, \hat{W}$ (later index $\wedge$ is suppressed)

$$\rho_0 \frac{\partial U_1}{\partial t} + \frac{\partial p}{\partial x} = Q_x + F_x = S_x$$

$$\rho_0 \frac{\partial U_2}{\partial t} + \frac{\partial p}{\partial y} = Q_y + F_y = S_y \qquad (4)$$

$$\rho_0 \frac{\partial W}{\partial t} + \frac{\partial p}{\partial z} + g\rho = Q_z + F_z = S_z$$

$$\rho_0 \left[ \frac{\partial U_1}{\partial x} + \frac{\partial U_2}{\partial y} + \frac{\partial W}{\partial t} \right] = R + M = \Phi$$

$$\frac{\partial \rho}{\partial t} + W \frac{\partial \rho_0}{\partial z} = T$$

Here the right-hand members $Q_x, Q_y, Q_z, R, T$ include members due to the medium rotation, viscosity, compressibility of the medium and nonlinearity of the medium

$$Q_x = -\rho \frac{\partial U_1}{\partial t} - (\rho + \rho_0)\left[ U_1 \frac{\partial U_1}{\partial x} + U_2 \frac{\partial U_1}{\partial y} + W \frac{\partial U_1}{\partial t} + f U_2 \right] + \nu\left[ \Delta_3 U + \frac{\partial D}{\partial x} \right]$$

$$Q_y = -\rho \frac{\partial U_2}{\partial t} - (\rho + \rho_0)\left[ U_1 \frac{\partial U_2}{\partial x} + U_2 \frac{\partial U_2}{\partial y} + W \frac{\partial U_2}{\partial t} - f U_1 \right] + \nu\left[ \Delta_3 V + \frac{\partial D}{\partial y} \right]$$

$$Q_z = -\rho \frac{\partial W}{\partial t} - (\rho + \rho_0)\left[ U_1 \frac{\partial W}{\partial x} + U_2 \frac{\partial W}{\partial y} + W \frac{\partial W}{\partial z} \right] + \nu\left[ \Delta_3 W + \frac{\partial D}{\partial z} \right]$$

(5)

$$R = -\frac{1}{c^2}\left[ \frac{\partial p}{\partial t} + U_1 \frac{\partial p}{\partial x} + U_2 \frac{\partial p}{\partial y} + W \frac{\partial p}{\partial z} \right] - \rho D$$

$$T = -U_1 \frac{\partial \rho}{\partial x} - U_2 \frac{\partial \rho}{\partial y} - W \frac{\partial \rho}{\partial z} + \frac{1}{c^2}\left[ \frac{\partial p}{\partial t} + U_1 \frac{\partial p}{\partial x} + U_2 \frac{\partial p}{\partial y} + W \frac{\partial p}{\partial z} \right]$$

$$D = \frac{\partial U_1}{\partial x} + \frac{\partial U_2}{\partial y} + \frac{\partial W}{\partial z}$$

The internal gravity waves have extremely small perturbations $\varsigma$ on the surface. Therefore while linearizing the boundary conditions on the surface $z = 0$, we obtain:

$$W\big|_{z=0} = \frac{\partial \varsigma}{\partial t}\bigg|_{z=0}$$

$$p(x, y, 0, t) = p(x, y, \varsigma, t) - \varsigma \frac{\partial p}{\partial z}$$

or by replacing $\frac{\partial p}{\partial z}$ with $\frac{\partial p_0}{\partial z} = -g\rho_0$, we have:

$$p(x, y, 0, t) = \varsigma(x, y, t) g \rho_0(0)$$

Whence by differentiating with respect to $t$, we obtain:

$$\frac{\partial p}{\partial t} - W g \rho_0(0) = 0 \bigg|_{z=0} \qquad (6)$$

By excluding from the combined equations (4) the variables $U, V, \rho, p$, we obtain for the velocity vertical component the internal gravitational waves ordinary equation with some non-zero right-hand member

$$\frac{\partial^2}{\partial t^2}\Delta_3 W + N^2 \Delta W = P \qquad (7)$$

$$N^2 = -\frac{g}{\rho_0}\frac{d\rho_0}{dz}$$

$$P = \frac{N^2(z)}{g}\frac{\partial^3 W}{\partial z \partial t^2} - \frac{1}{\rho_0}\frac{\partial^2}{\partial z \partial t}\left[\frac{\partial S_x}{\partial x} + \frac{\partial S_y}{\partial y} - \frac{\partial \Phi}{\partial t}\right] + \frac{\Delta_h}{\rho_0}\frac{\partial S_z}{\partial t} - \frac{g}{\rho_0}\Delta T \qquad (8)$$

The boundary condition at the bottom $z = -H$ maintains the form (2A) ir (2B). By operating $\Delta$ on the formula (6) and expressing $\Delta p_t$ in terms of $W$, we obtain the condition with $z = 0$

$$\frac{\partial^3 W}{\partial z \partial t^2} - g\Delta W = \frac{1}{\rho_0}\left[\frac{\partial^2 \Phi}{\partial t^2} - \frac{\partial^2 S_x}{\partial t \partial x} - \frac{\partial^2 S_y}{\partial t \partial y}\right] \qquad (9)$$

In the equations (4) and the boundary condition (9) the right-hand members are the sum of two summands depending on the extraneous sources (of body forces $\vec{F}$ and mass sources density $M$) and on viscous small corrections, $\nu$, on compressibility of the medium (i.e. nonzero value $c^{-2}$), on medium rotation $f$, as well as on the corrections due to the Boussinesq approximation and being of order $\frac{N^2}{g} \ll 1$.

The equation (7) is convenient for further evaluation with these corrections perturbation [1]. Therefore internal gravity waves propagation with small additives due to for example nonlinear terms is described by the equation

$$\frac{\partial^2}{\partial t^2}\Delta_3 W + N^2 \Delta W = \varepsilon P(W) \tag{10}$$

where in the formula for $P$ only (10) only summands considering nonlinearity of initial combined hydrodynamic equations will be included, and the parameter $\varepsilon$ takes into account these summands are small. Later formal expansion of the equation solution (10) in power of $\varepsilon$ is used. Let us consider $\varepsilon P$ correction influence on internal gravitational wave one separate wave mode propagation. For the purpose of calculations simplification a horizontal case is considered, i.e. there's no y-dependence, i.e. $U_2 \equiv 0$. Let us later denote $U_1 \equiv U$, then the problem can be set the following way. Let us assume that the correction $\varepsilon P$ is included with $t = 0$

$$\frac{\partial^2}{\partial t^2}\left(\frac{\partial^2 W}{\partial z^2} + \frac{\partial^2 W}{\partial x^2}\right) + N^2(z)\frac{\partial^2 W}{\partial x^2} = \Theta(t)\varepsilon P(W) \tag{11}$$

where $\Theta(t) = 0$ with $t < 0$ and $\Theta(t) = 1$ with $t > 0$, and the right-hand member $P$ of this equation is given by

$$P = \frac{\partial^2}{\partial x \partial t}\left(W(\frac{\partial^2 U}{\partial z^2} + \frac{\partial^2 U}{\partial x^2}) - U(\frac{\partial^2 W}{\partial z^2} + \frac{\partial^2 W}{\partial x^2}) + \rho(\frac{\partial^2 U}{\partial t \partial z} - \frac{\partial^2 W}{\partial t \partial x}) + \frac{\partial \rho}{\partial z}\frac{\partial U}{\partial t} - \frac{\partial \rho}{\partial x}\frac{\partial W}{\partial t}\right) +$$

$$+ g \frac{\partial^2}{\partial x^2}(U \frac{\partial \rho}{\partial x} + W \frac{\partial \rho}{\partial z})$$

where $\rho$ is the density perturbation normalized to a some typical value of nonperturbed density $\rho_0^*$. With $t < 0$ the equation solution (11) is used in form of eigen wave mode.

$$W_0 = A \varphi_n(z,k) \cos(\omega_n(k)t - kx) \qquad (12)$$

where $\omega_n(k)$ andи $\varphi_n(z,k)$ are dispersion curves and normalized eigenfunctions of primal vertical spectral problem of the internal gravity waves accordingly [1-5]

$$\frac{\partial^2 \varphi_n(z,k)}{\partial z^2} + k^2 (\frac{N^2(z)}{\omega_n^2(k)} - 1) \varphi_n(z,k) = 0$$

(13)

$$\varphi_n(z,k) = 0, z = 0, -H$$

Then corresponding to the null approximation $W_0$ horizontal velocity $U_0$ and density perturbation $\rho_0$ are given by

$$U_0 = \frac{A}{k} \frac{\partial \varphi_n(z,k)}{\partial z} \sin(\omega_n(k)t - kx)$$

(14)

$$\rho_0 = \frac{A N^2(z)}{\omega_n(k) g} \varphi_n(z,k) \sin(\omega_n(k)t - kx)$$

With $t > 0$ we will seek the solution (11) in form of series in terms of powers of small paramater $\varepsilon$

$$W = W_0 + \varepsilon W_1 + \varepsilon^2 W_2 + ...$$

It is evident that for function $W_1$ definition the equation is obtained

$$\frac{\partial^2}{\partial t^2}\left(\frac{\partial^2 W_1}{\partial z^2} + \frac{\partial^2 W_1}{\partial x^2}\right) + N^2(z)\frac{\partial^2 W_1}{\partial x^2} = P(W_0) \tag{15}$$

The equation (11) solution coinciding with $W_0$ with $t < 0$ is continuum-level with its derivative with respect to $t$. Therefore with $t = 0$ the function $W_1$ and its derivative with respect to $t$ go to zero.

$$W_1 = \frac{\partial W_1}{\partial t} = 0 \tag{16}$$

As in actual ocean practice as a rule only first modes [6-9] are excited, later we will for definiteness consider in (12) the first mode. Taking account of (12)-(14) the right-hand member in (15) is given by

$$P(W_0) = A^2 2k \sin(2\omega_1(k)t - 2kx)\left[\frac{N^2(z)\omega_1(k)}{gk}\left(\varphi_1(z,k)\frac{\partial^2 \varphi_1(z,k)}{\partial z^2} + (\frac{\partial \varphi_1(z,k)}{\partial z})^2\right) - \right.$$

$$-\frac{2N^2(z)k\omega_1(k)(\varphi_1(z,k))^2}{g} - \frac{\omega_1(k)}{k}\left(\frac{\partial \varphi_1(z,k)}{\partial z}\frac{\partial^2 \varphi_1(z,k)}{\partial z^2} - \varphi_1(z,k)\frac{\partial^3 \varphi_1(z,k)}{\partial z^3}\right) +$$

$$+ \frac{\omega_1(k)\varphi_1(z,k)}{gk}\frac{\partial \varphi_1(z,k)}{\partial z}\frac{\partial N^2(z)}{\partial z} - \frac{k}{\omega_1(k)}\left(\frac{\partial \varphi_1(z,k)}{\partial z}\right)^2 \frac{\partial N^2(z)}{\partial z}\Bigg] =$$

$$\equiv \Phi(z,k)\sin(2\omega_1(k)t - 2kx) \qquad (17)$$

We will seek the equation (15) solution in form of problem (13) eigen-function series

$$W_1 = \sin(2\omega_1(k)t - 2kx)\sum_{i=1}^{\infty} d_i \varphi_i(z,2k) \qquad (18)$$

We will conceive the right-hand member (17) of the equation (15) similar of as

$$\Phi(z,k) = N^2(z)\sum_{i=1}^{\infty} c_i \varphi_i(z,2k), \quad c_i = \int_{-H}^{0} \Phi(z,k)\varphi_i(z,2k)dz \qquad (19)$$

Substituting (18) and (19) in (15) we obtain

$$d_i = \frac{c_i \omega_i^2(2k)}{4k^2(4\omega_1^2(k) - \omega_i^2(2k))}$$

Taking into account the initial conditions (16), we have

$$W_1 = \sum_{i=1}^{\infty} d_i \varphi_i(z,2k) \left[ \sin(2\omega_1(k)t - kx) - \frac{\omega_i(2k)\sin^2(2kx) + 2\omega_1(k)\cos^2(2kx)}{\omega_i(2k)} \times \right.$$

(20)

$$\left. \times \sin(2\omega_1(k)t - 2kx) - \frac{\sin(4kx)}{2\omega_i(2k)}(2\omega_1(k) - \omega_i(2k))\cos(2\omega_1(k)t - 2kx) \right]$$

The (20) indicates, that the greatest contribution to $W_1$ is made by the summand with the multiplier $d_1$. Our aim is to compare the correction $W_1$ with the unperturbed solution $W_0$. Let us substitute the first summand in (20) with a secular (resonance) term, (assuming $\omega_1(2k) = 2\omega_1(k)$) thus only extending $W_1$

$$\widetilde{W}_1 = \frac{a_1 \omega_1(2k)}{8k^2} \varphi_1(z,2k) t \cos(\omega_1(2k)t - 2kx) \qquad (21)$$

and let us evaluate time, within which $\widetilde{W}_1$ begins to be comparable with $W_0$.

For coefficient $a_1$ calculation we assume $\varphi_1(z,k) \approx \varphi_1(z,2k)$, which is correct for sufficiently small $k$. As a result we obtain

$$a_1 = 2A^2 \left( -\frac{\omega_1(k)}{g} \int_{-H}^{0} N^2(z)\varphi_1(z,k)\left(\frac{\partial \varphi_1(z,k)}{\partial z}\right)^2 dz - \frac{2k^2\omega_1(k)}{g} \int_{-H}^{0} N^2(z)(\varphi_1(z,k))^3 dz + \right.$$

(22)

$$+\frac{6k^2}{\omega_1(k)}\int_{-H}^{0} N^2(z)(\varphi_1(z,k))^2 \frac{\partial \varphi_1(z,k)}{\partial z} \, dz - 3\omega_1(k)k^2 \int_{-H}^{0} (\varphi_1(z,k))^2 \frac{\partial \varphi_1(z,k)}{\partial z} \, dz \Bigg)$$

To use formulae (21) and (22) let us consider fluid layer $H = 100$ m deep, $N(z) = const = 0.01 c^{-1}$, then $\varphi_1 = \frac{\sqrt{2}}{N\sqrt{H}} \sin \frac{\pi z}{H}$. Later let us take $k = 0.02$, then the wavelength $\lambda \sim 300$ m, and $A = 10^{-4} м^{3/2} c^{-2}$, which correspond with undisturbed wave amplitude in (14) approximately 0,0014 m/s. For the case $N(z) = const$ the latest two summands in (22) are zero, the first two integrals are easily calculated and then $a_1 = -1.4 \cdot 10^{-13}$. Let us evaluate time, within which $\widetilde{W}_1$ are in the region of 5% of the value $W_0$: $t = \frac{4k^2}{c_1 \omega_1(2k)} A \cdot 0,05 = 1,5 \cdot 10^7$ seconds, which far exceeds typical oscillatory periods studied in [1, 6-9] of wave processes. Thus in interesting for us wavelength range we can use linear approximation during study of internal gravity waves dynamics. Similarly it is easy to evaluate also influence of other corrections to the linear theory of internal gravity waves generation and propagation, and the results of these evaluation indicate adequacy and supportability of used in [1-5] model wave dynamics.